\documentclass[letterpaper, 10 pt, conference]{ieeeconf}  

\IEEEoverridecommandlockouts                              

\usepackage{graphics} 
\usepackage{epsfig} 
\usepackage[export]{adjustbox}
\usepackage{times} 
\usepackage{algorithm, color}
\usepackage{algpseudocode,algorithmicx}
\usepackage{subfigure}
\usepackage{amsmath} 
\usepackage{amssymb}  
\usepackage{hyperref}
\newtheorem{proposition}{Proposition}
\newtheorem{theorem}{Theorem}

\newtheorem{lemma}{Lemma}

\newtheorem{remark}{Remark}

\newtheorem{example}{Example}

\usepackage{scalerel,stackengine}
\stackMath
\newcommand\reallywidehat[1]{%
\savestack{\tmpbox}{\stretchto{%
  \scaleto{%
    \scalerel*[\widthof{\ensuremath{#1}}]{\kern-.6pt\bigwedge\kern-.6pt}%
    {\rule[-\textheight/2]{1ex}{\textheight}}
  }{\textheight}%
}{0.5ex}}%
\stackon[1pt]{#1}{\tmpbox}%
}

\title{\LARGE \bf
Neural Exponential Stabilization of  Control-affine Nonlinear Systems
}

\author{Muhammad Zakwan$^{1}$, Liang Xu$^{2}$, and Giancarlo Ferrari-Trecate$^{1}$
\thanks{This research is supported by the Swiss National Science Foundation under the NCCR Automation (grant agreement 51NF40 180545), and the National Natural Science Foundation of China under Grant 62373239, 62333011.}
\thanks{Muhammad Zakwan and Giancarlo Ferrari-Trecate are with the Institute of Mechanical Engineering, Ecole Polytechnique Fédérale de Lausanne (EPFL), CH-1015 Lausanne, Switzerland,  email: \{muhammad.zakwan, giancarlo.ferraritrecate\}@epfl.ch;
Liang Xu is with the School of Future Technology, Shanghai University, Shanghai, China, email: liang-xu@shu.edu.cn}%
}

\begin{document}

\maketitle
\thispagestyle{empty}
\pagestyle{empty}

\begin{abstract}
This paper proposes a novel learning-based approach for achieving exponential stabilization of nonlinear control-affine systems. 
We leverage the Control Contraction Metrics (CCMs) framework to co-synthesize Neural Contraction Metrics (NCMs) and Neural Network (NN) controllers. 
First, we transform the infinite-dimensional semi-definite program (SDP) for CCM computation into a tractable inequality feasibility problem using element-wise bounds of matrix-valued functions. 
The terms in the inequality can be efficiently computed by our novel algorithms. Second, we propose a free parametrization of NCMs guaranteeing positive definiteness and the satisfaction of a partial differential equation, regardless of trainable parameters. Third, this parametrization and the inequality condition enable the design of contractivity-enforcing regularizers, which can be incorporated while designing the NN controller for exponential stabilization of the underlying nonlinear systems. Furthermore, when the training loss goes to zero, we provide formal guarantees on verification of the NCM and the exponentional stabilization under the NN controller. Finally, we validate our method through benchmark experiments on set-point stabilization and increasing the region of attraction of a locally pre-stabilized closed-loop system. 
\end{abstract}


\section{INTRODUCTION}

Learning-enabled control has demonstrated state-of-the-art empirical performance on various challenging tasks in robotics \cite{xiao2023barriernet}. However, this performance boost often comes at the cost of safety and stability guarantees, and robustness to disturbances of the learned NN controllers \cite{sun2021learning}. Traditionally, engineers faced challenges in manually designing certificates such as Lyapunov functions \cite{taylor2019episodic}, barrier functions \cite{peruffo2021automated}, and contraction metrics \cite{tsukamoto2021contraction}, often relying on intuition or experience for specific applications. While numerical methods like sum-of-squares (SoS) have emerged for computing these certificates, many remain impractical \cite{sun2021learning}. To address these limitations, researchers have employed Neural Networks (NNs) to learn both control policies and safety functions, referred to as neural certificates  \cite{chang2019neural,richards2018lyapunov}. 
These methods have been applied successfully to complex nonlinear control tasks such as stable walking \cite{choi2020reinforcement}, quadrotor flight \cite{sun2021learning}, and safe decentralized control of multi-agent systems \cite{qin2021learning}.  Besides, designing NN controller on dynamical system, several  approaches propose 
 modelling the system under control as a NN from data  \cite{armenio2019model,bonassi2022recurrent, terzi2021learning} and later ensure the stability of the closed-loop system.
 However, the main challenge lies in finding tractable and scalable verification techniques for these neural certificates.

In this paper, we train a state feedback controller, parametrized by slope-restricted neural networks (NNs), to achieve exponential stabilization of control-affine nonlinear systems. To this end, we leverage the well-established framework of Control Contraction Metrics (CCMs) \cite{tsukamoto2021contraction,manchester2017control}. While the existence of CCMs for a dynamical system ensures exponential stabilization, they usually leverage infinite-dimensional  intractable inequalities. To tackle this,  we derive a tractable inequality guaranteeing exponential stabilizability of the closed-loop system. 
This condition relies on element-wise bounds of some specific matrix-valued functions. Moreover, we parametrize the CCMs via NNs and guarantee that they are inherently positive definite and satisfy a PDE for all trainable parameters. We call them Neural Contraction Metrics (NCMs). By leveraging both the stabilizability inequality and NCMs, we design regularizers that enforce closed-loop contractivity. These regularizers directly penalize the NN controller's weights based on the Gershgorin disk theorem.
Moreover, if the regularizer loss goes to zero, we can formally guarantee the verification of the learned NCM certificate for the NN controller.

{\bf Related works:} Traditional methods for formulating CCMs, such as SoS programming \cite{tan2008stability} and Reproducing Kernal Hilbert Space (RKHS) theory \cite{chang2019neural}, suffer from limitations. These include structural restrictions on control input matrices, separate synthesis of controllers and CCMs, assumptions about system dynamics (polynomial or approximated as such), and poor scalability \cite{dawson2023safe}. These drawbacks make these approaches impractical for many real-world applications. Other approximation techniques using gridding methods lack rigorous guarantees \cite{manchester2017control}.
Our proposed method addresses some of the limitations of existing approaches. It provides strong guarantees when the loss function approaches zero and scales well to handle a large number of states. However, there is still room for improvement  in terms of conservatism.

Recent research explores synthesizing safety certificates through NNs. Both the controller and certificates are usually parameterized via NNs, followed by verification \cite{dawson2023safe}. 
A popular technique, the learner-verifier approach (also called counter-example guided inductive synthesis), involves training a certificate network while a verifier (usually using a satisfiability modulo theory (SMT) solver) assesses its feasibility \cite{dawson2023safe}. If valid, the verifier halts training; if invalid, it provides counterexamples for training data enrichment.
Verification can be implemented in various ways. For instance, if given piecewise affine dynamics and NNs (for controllers and certificates) are implemented with ReLU activations, verification can be formulated as a Mixed-Integer Linear Program (MILP) solved with standard solvers \cite{dai2020counter,schwan2023stability}. 
These methods have gained traction and have been applied to learning Lyapunov functions for general nonlinear systems  \cite{chang2019neural, abate2020formal} and hybrid systems \cite{chen2021learning}.
However, both SMT- (NP-complete \cite{dawson2023safe}) and MILP-based methods are computationally expensive and their complexity grows exponentially with the number of neurons in the NN certificate, limiting their applicability \cite{dawson2023safe}. 
Additionally, Lyapunov-based methods are restricted to non-autonomous systems. On the contrary, CCMs, being more general, can handle time-varying systems and are the focus of this paper.

Some recent works explore learning CCMs with NNs. For example, the work \cite{tsukamoto2020neural} uses recurrent NNs to parametrize the contraction metric before constructing the controller. However, this approach assumes specific dynamics and is limited to convex combinations of state-dependent coefficients. In contrast,  under mild assumptions, our method can simultaneously synthesize the controller and the associated CCM certificate for general control-affine systems.
The work \cite{sun2021learning} also proposes an NN-based framework for co-synthesizing CCMs and controllers for control-affine systems. However, there are several key differences with our work. 
First, \cite{sun2021learning} addresses tracking problems, assuming access to desired trajectories and control policies. We focus instead on the exponential stabilization of equilibrium, a different problem formulation. 
Second, we provide a free-parametrization of NCMs that are positive definite by design and satisfy a PDE constraint characteristic of CCMs for all trainable parameters. On the other hand, the authors of \cite{sun2021learning} try to penalize this constraint empirically, and do not provide rigorous guarantees. Finally, in specific cases, for instance, when the underlying system is global Lipschitz, our method  provides global stability guarantees, which is not the case in \cite{sun2021learning}.


{\bf Contributions:} The contributions of our paper can be summarized as follows:
\begin{itemize}
    \item We propose an NN training problem that simultaneously learns both the controller and the CCM for achieving exponential stabilization of control-affine nonlinear systems. We provide formal guarantees for the verification of the learned NCM under the condition that the training loss converges to zero. Furthermore, our approach is scalable w.r.t both  the dimension of state and the number of neurons in the NN controller.
    
    \item We introduce a novel free parametrization for NCMs. These NCMs are inherently positive definite, and their Jacobians satisfy a PDE constraint regardless of the trainable parameters. This eliminates the need for explicit penalization of the constraint violation during training, simplifying the training process.
    
    \item We develop an algorithm that computes element-wise upper and lower bounds of the Jacobian of the NN controller. These bounds are then leveraged to transform the verification of CCMs from an infinite-dimensional SDP into a more tractable finite-dimensional inequality (see Theorem~\ref{thm2}).
    
    \item We validate the efficacy of our method through benchmark experiments on two key tasks: set-point stabilization and expanding the region of attraction of a locally pre-stabilized closed-loop system.
\end{itemize}

{\bf Organization:} Section~\ref{sec:prelim} provides a brief overview of contraction theory and CCMs. Section~\ref{sec:results} presents our core contributions: designing regularizers that enforce closed-loop contractivity, introducing a free parametrization of NCMs, and outlining Algorithm~\ref{alg} to facilitate implementation of these regularizers. Section~\ref{sec:experiments} validates the proposed methods through benchmark experiments. Finally, Section~\ref{sec:conclusions} draws some conclusions.

{\bf Notation:} We denote by $\mathbb{R}$ and $\mathbb{R}_{\geq 0}$ the set of real and non-negative real numbers respectively. For a symmetric matrix $A \in \mathbb{R}^{n \times n}$, the notation $A \succ 0 (A \prec 0)$ means $A$ is positive (negative) definite. The set of positive definite $n \times n$ matrices is denoted by $\mathbb{S}_{\succ 0}$. For a matrix-valued function $M(x): \mathbb{R}^n \mapsto \mathbb{R}^{n \times n}$, its element-wise Lie derivative along a vector $v \in  \mathbb{R}^n$ is $ \partial_v M:= \sum_i v_i \frac{\partial M}{\partial x_i}$.  Unless
otherwise stated, $x_i$ denotes the i-th element of vector $x$. For a matrix $M \in \mathbb{R}^{n \times n}$, we denote $M + M^\top$ by $\text{Sym}[M]$. For two matrices $A$ and $B$ of the same dimensions, the notation $A < (\leq) B$ means that all the entries of $A$ are element-wise less (less-equal) than the entries of $B$. The maximum eigenvalue of a symmetric matrix   $A$ is given by $\bar{\lambda}(A)$. The kernel of a linear map $g$ is defined by $\textbf{Ker}(g)$.

\section{Preliminaries} \label{sec:prelim}
In this paper, we consider control-affine systems as
\begin{align} \label{eq:plant}
    \dot{x} = f(x) + gu(t) \;,
\end{align}
where $x  \in  \mathcal{X} \subseteq \mathbb{R}^n$, $u(t) \in \mathcal{U} \subseteq \mathbb{R}^m$ for all $t \in \mathbb{R}_{\geq 0}$ are states, and inputs respectively. Here, $\mathcal{X}$, and $\mathcal{U}$  are compact state and input sets, respectively. We assume that $f: \mathbb{R}^n \mapsto \mathbb{R}^n$ is a  smooth map, the control input $u: \mathbb{R}_{\geq 0}\mapsto \mathcal{U}$ is a piece-wise continuous function. The goal of this paper is to design a  NN state-feedback controller $u_\theta(x)$, where $\theta$ are the trainable parameters, such that the controlled trajectory $x(t)$ can reach the desired equilibrium point  $x^\star$ whenever $x(0)$ is in a neighborhood of $x^\star$. 
In this work, we leverage contraction theory to achieve our goal.

Contraction theory  \cite{lohmiller1998contraction, tsukamoto2021contraction} analyzes the incremental stability of systems by examining the evolution of the distance between neighboring trajectories. For a time-invariant autonomous system $\dot{x} = f (x)$,
given a pair of neighboring trajectories denote the infinitesimal displacement between them by $\delta x$ (also called a virtual displacement). The evolution of $\delta x$ can be represented by a linear time-varying (LTV) system: $\dot{\delta} x = \partial_x f(x) \delta x$.
Then, the squared distance between these trajectories $\delta x^\top \delta x$, evolves according to  $ \frac{d}{dt} (\delta x^\top \delta x)
= 2 \delta x \dot{\delta}x= 2 \delta x^\top \partial_x f(x) \delta x $.
If the symmetric part of the Jacobian $\partial_x f$ is uniformly negative definite, i.e.,  $ \frac{1}{2} ( \partial_x f + \partial_x f^\top ) \preceq -\rho I$ for some $\rho > 0$, then the system is contracting.
This condition ensures that $\delta x^\top \delta x$ converges exponentially to zero at a rate of $2 \rho$.
Consequently, all trajectories of the system converge to a common equilibrium trajectory \cite{lohmiller1998contraction}. 

The concept of contraction can be generalized using a contraction metric $M: \mathbb{R}^n \mapsto \mathbb{S}_{\succ 0}$, which is a smooth matrix-valued function.
Since $M(x)$ is always positive definite, if $\delta x^\top M(x) \delta x$ converges exponentially to zero, the system is contracting. The converse is also true \cite{manchester2017control}.
Contraction theory can be further extended to control-affine systems. First, the corresponding differential dynamics of \eqref{eq:plant} is given by
\begin{equation*}
    \dot{\delta} x = F(x) \delta x + g \delta u \;,
\end{equation*}
where $F(x) := \partial_x f $, and $\delta u$ is the infinitesimal displacement between two neighboring inputs.
A fundamental theorem in Control Contraction Metric (CCM) theory \cite{manchester2017control} says that if there exists a metric $M (x)$ such that the following
conditions hold for all $x$ and some $\rho > 0$
  \begin{align*}
\dot{M}(x) +  \text{Sym}[M(x) \partial_x(f(x)+gu)] \prec - 2 \rho M(x) \;.
  \end{align*}
Then, the closed-loop system is contracting with rate $\rho$ under metric $M(x)$ \cite{manchester2017control}.
However, as pointed out in \cite{manchester2017control}, finding a suitable metric or designing a controller based on a predetermined metric can be challenging. To address this limitation, we propose a method that leverages NNs to learn both the metric and the controller simultaneously, as explored in \cite{sun2021learning}.




\section{Main Results} \label{sec:results}

This section leverages contraction theory to design NN controllers for exponential stabilization. We begin by introducing a class of NNs with slope-restricted activation functions, well-suited for parameterizing the control policy for the system (\ref{eq:plant}). Next, we establish a sufficient condition to guarantee the exponential stabilization of the  system (\ref{eq:plant}). This condition relies on element-wise upper and lower bounds of terms crucial for CCM synthesis. We also provide algorithms for computing these bounds efficiently. Furthermore, we propose a novel free parametrization of NCMs that ensures they are inherently positive definite and their Jacobians  satisfy CCM-related PDE constraint by design. By leveraging our proposed condition, and NCMs, we design regularizers that enforce contractivity in the closed-loop system. 


{\bf NN architecture:} We consider the following class of NN controllers 
\begin{align}
    z_1 &= \sigma_{1} \left ( W_{1}x+b_{1} \right ) \nonumber \\ 
    z_\ell &= \sigma_{\ell} \left ( W_{\ell}z _{\ell -1}+b_{\ell} \right ), \ \ell = 2, \cdots, N \;,      \label{eq:controller} \\ 
    u_\theta &= W_o z_{N}  \;,\nonumber
\end{align}  
where $N \geq 1$ represents the depth of the NN, and the set $\theta := \{W_o, W_N, \cdots, W_1, b_1,\cdots, b_N \}$ contains all the trainable parameters.
We consider a class of slope-restricted activation functions $\sigma(\cdot)$ such that their derivative w.r.t. the input satisfies $0 < a \leq \sigma'(\cdot) < b$.
A typical activation function that satisfies this assumption is the smooth LeakyReLu
\begin{align} \label{eq:activation}
    \sigma(x) = \alpha x+ (1 - \alpha) \log(1+e^x) \;,
\end{align}
where $\alpha$ controls the angle of the negative slope.
The proposed parametrization exhibits universal approximation properties \cite{pmlr-v202-li23g}.



{\bf Neural contractive closed-loop system:} Before presenting our main results, let us define the set containing the desired equilibrium point $x^\star$ for the system \eqref{eq:plant} as
  \begin{align*}
      \mathcal{E} := \bigg\{ x \in \mathcal{X} | f(x^\star) + gu_\theta(x^\star) = 0 \bigg\} \;.
  \end{align*}
Then, the following result holds, adapted from \cite{manchester2017control}
\begin{theorem} \label{thm1}
  Suppose the set $\mathcal{E}$ is non-empty, and there exists a NN $M_\phi: \mathcal{X} \mapsto \mathbb{S}_{\succ 0}$ endowed with some trainable parameters $\phi$, such that 
    \begin{gather} \label{eq:cont_ineq}
     \dot{M}_\phi+ \text{Sym} \left[ M_\phi  \partial_x (f(x) + gu_\theta(x)) \right] \prec  -  2\rho M_\phi \;
    \end{gather}
    holds for all $x \in \mathcal{X}$ and for some $\rho >0$.
    Then, the equilibrium point $x^\star$ is exponentially stable. \hfill $\blacksquare$
  \end{theorem}

Note that the inequality \eqref{eq:cont_ineq} in Theorem \ref{thm1} is an infinite-dimensional SDP and can be cumbersome to verify for all $x \in \mathcal{X}$ using standard convex optimization solvers \cite{manchester2017control}.
To tackle this issue, we provide a sufficient condition for ensuring the exponential stability of the closed-loop system. Specifically, this condition is based on the Gershgorin disc theorem and the proof is provided in Appendix.

\begin{theorem} \label{thm2}
Suppose $x^\star \in \mathcal{E}$ is  the desired equilibrium point of the system \eqref{eq:plant} and there exists a metric NCM $M_\phi: \mathcal{X} \mapsto \mathbb{S}_{\succ 0}$ endowed with some trainable parameters $\phi$. Then, $x^\star$ is exponentially stable if the following inequality is satisfied
\begin{align} \label{eq:thm2_main}
     2U_{ii}+ \eta \le - \sum_{j\neq i} \max \left\{ |L_{ij}+L_{ji}|, |U_{ij}+U_{ji}|\right\} \;,
\end{align}
where $U_{ii}$ is the $i$th diagonal entry of the matrix $U = U_\theta + U_\phi$, and $L_{ij}$ is the $(i,j)$ entry of the matrix $L = L_\theta + L_\phi$ and $L_\theta, L_\phi, U_\theta, U_\phi$ verify
\begin{align*}
     L_\theta \leq  M_\phi(x) g \partial_x u_\theta(x) \leq U_\theta, \  
     L_\phi \leq \dot{M}_\phi(x) \leq U_\phi \;.
\end{align*}
Moreover, $\eta = - (c_1 + c_2 )$ is a scalar and  the constants $c_1$ and $c_2$ satisfy 
\begin{align*}
c_1 \geq 2 \rho \sup_{s \in \mathcal{X}} \bar{\lambda}(M_\phi(s)),  
    c_2 \geq \sup_{s \in \mathcal{X}}\bar{\lambda}\left(\text{Sym}[M_\phi(s) \partial_s f(s)]  \right) 
\end{align*}
for all $x \in \mathcal{X}$. \hfill $\blacksquare$
\end{theorem}

Once the constants $c_1$ and $c_2$, and the element-wise bounds $L$ and $U$ have been determined, condition (\ref{eq:thm2_main}) becomes a finite-dimensional scalar inequality, making its verification significantly easier. While Theorem~\ref{thm2} leverages the Gershgorin disc theorem, resulting in a more conservative condition compared to Theorem~\ref{thm1}, this trade-off is necessary for tractability. Nevertheless, as shown in Section~\ref{sec:experiments}, good performance can still be achieved despite this conservatism.

{\bf Neural Contraction Metric (NCM):} An integral part of Theorem \ref{thm2} is the parametrization of the CCM by a NN. 
In this paper, motivated by \cite{sun2021learning}, we parametrize NCM as $ M_\phi(x) = \Gamma_\phi(x)^\top \Gamma_\phi(x) + \epsilon I$, where  $\Gamma_\phi: \mathcal{X} \mapsto \mathbb{R}^{n \times n}$ is a smooth matrix-valued function depending on some trainable weights $\phi$, and $\epsilon >0$ is a small positive constant. Since $M_\phi(x)$ is positive definite irrespective of the weight matrices $\phi$ and for all values of $x \in \mathcal{X}$, as desired, we call it a free parametrization. Besides the positive definiteness of $M_\phi(x)$, in  inequality \eqref{eq:thm2_main}, $\dot{M}_\phi(x)$ is a matrix with $(i,j)$ elements given by $\partial_x \left( M_{\phi,ij}^\top(x) \right) (f(x) + gu_\theta(x))$, which  are affine in the control signal $u$. Therefore, for inequality \eqref{eq:thm2_main} to hold for all $u$, it is crucial that
\begin{align} \label{eq:pde}
    \partial_x \left( M_{\phi,ij}^\top(x) \right) g = 0  
\end{align}
for all $x \in \mathcal{X}$. The constraint \eqref{eq:pde} is a PDE and is cumbersome to solve. 
Motivated by \cite{xu2022neural}, the next result provides a parametrization of the NCM $M_\phi(x)$ such that the state-dependent equality \eqref{eq:pde} is satisfied by design, that is, regardless of the choice of trainable parameters.
\begin{proposition} \label{prop:ccm}
   Suppose the dimension of $\textbf{Ker}(g)$ is $r$, and let $v_1,...,v_r$ be the basis of $\textbf{Ker}(g)$. Define each entry of $\Gamma_{\phi}(x)$ as 
   \begin{align} \label{eq:param}
 \Gamma_{\phi,ij}(x) := K_{ij}\left( \sum_{\ell =1}^r \beta_{\ell,ij} (x^\top v_\ell)\right), \forall \  i,j \;,
   \end{align}
 for some continuously differentiable NNs $K_{ij}: \mathbb{R}\mapsto \mathbb{R}, \beta_{\ell,ij}:\mathbb{R}\mapsto \mathbb{R}, \ell=1, \ldots, r$. Then, taking  $ M_\phi(x) = \Gamma_\phi(x)^\top \Gamma_\phi(x) + \epsilon I$ satisfies \eqref{eq:pde} for all $x \in \mathcal{X}.$ \hfill $\blacksquare$
\end{proposition}

The proof is provided in the Appendix. Notably, in most real-world applications, the systems are under-actuated where the rank of the control matrix, i.e. input dimension is lower than the number of states ($m<n$). This guarantees the existence of a non-trivial subspace $\textbf{Ker}(g)$. The following example showcases the essence of Proposition~\ref{prop:ccm} by demonstrating how to parameterize NCMs for a single-input control-affine system.
 
\begin{example}[PDE \eqref{eq:pde} with $m=1$] Consider the system \eqref{eq:plant} with $m=1$, then one can parametrize each entry $(i,j)$ of $\Gamma_{\phi}(x)$ as 
\begin{align*}
    \Gamma_{\phi, ij}(x) = \gamma_{ij} \log (\cosh{(\alpha_{ij}g_\perp^\top x + b_{ij} )}) \;, 
\end{align*}
where $\gamma_{ij}, \alpha_{ij}, b_{ij}$ are scalar trainable parameters and $g_\perp$ is the left-annihilator of $g$ such that $g_\perp^\top g = 0$. Note that each entry of $M_{\phi,ij}(x)$ is given as 
\begin{align*}
    M_{\phi,ij}(x) = \sum_{k} \Gamma_{\phi, ik}^\top(x) \Gamma_{\phi, kj}(x)
\end{align*}
 for which we can analytically compute the Jacobian of both sides as follows:
 \begin{align*}
    \sum_{k} \partial_x (\Gamma_{\phi,ik}^\top& \Gamma_{\phi,k\ell})   = \sum_k \left( \Gamma_{\phi, ik}^\top \partial_x \Gamma_{\phi, kl} + \partial_x \Gamma_{\phi, ik}^\top \Gamma_{\phi, kl} \right)\\
   &= \sum_k \left(\Gamma_{\phi,ik}^\top \gamma_{kl} \alpha_{kl} \tanh (\alpha_{kl}g_\perp^\top x + b_{kl} )g_\perp  \right.\\
& \left. +  \Gamma_{\phi,kl}\gamma_{ik} \alpha_{ik} \tanh (\alpha_{ik}g_\perp^\top x + b_{ik})    g_\perp  \right)\\ 
\partial_x \left( M_{\phi,ij}^\top(x) \right) &=: h_\phi(x) g_\perp^\top \;, 
 \end{align*}
 where $h_\phi(x)$ is a scalar function. Replacing the above expression in \eqref{eq:pde} satisfies the PDE by design. \hfill $\blacksquare$
\end{example}
{\bf NN controllers with bounded Jacobians:}
To compute the element-wise lower and upper bounds $L_\theta$ and $U_\theta$ appearing in Theorem \ref{thm2}, it is crucial to calculate the element-wise bounds of the Jacobian of the NN controller \eqref{eq:controller}. To this end, we write the  Jacobian as:
\begin{equation} \label{eq:Jacobian}
    \partial_x u_\theta(x) =W_o J_NW_N J_{N-1} W_{N-1}\ldots J_1W_1 \;,
\end{equation}
where each $J_i = \sigma'(W_i x+ b_i)$ and satisfies $0 < a I \leq J_i \leq b I$.
Leveraging the special structure of the proposed NN controller \eqref{eq:controller}, in the following, we propose Algorithm \ref{alg} to compute element-wise lower $ L_{\partial_x u}$ and upper $U_{\partial_xu}$ bounds  of \eqref{eq:Jacobian}. 
\begin{algorithm} 
\begin{algorithmic}
\State    Set $L=I$, $U=I$
 \For{$i=1$, $\cdots$, $N$}
 \State call \textbf{Func1}($W_i, L, U$) and return $\tilde{L}, \tilde{U}$ (which calculates the element-wise upper and lower bound of $W_i J_{i-1} W_{i-1}\ldots J_1W_1$)
 \State call \textbf{Func2}($\tilde{L}, \tilde{U}, a, b)$ and return $\hat{L}, \hat{U}$ (which calculates the element-wise upper and lower bound of $J_iW_i J_{i-1} W_{i-1}\ldots J_1W_1$)
 \State set $L=\hat{L}, U=\hat{U}$
 \EndFor
 \State return $L, U$
 \end{algorithmic}  
 \caption{Returns the upper $U_{\partial_x u}$ and lower $L_{\partial_x u}$ bound of $\partial_x u_\theta$}
 \title{Algorithm 1}
 \label{alg}
\end{algorithm}
This algorithm relies on two functions: 
\begin{itemize}
  \item[$\bullet$] \textbf{Func1:} This function, provided in Appendix, calculates the element-wise lower $\tilde{L}$ and upper bound $\tilde{U}$ of a matrix product $WQ$. It takes the element-wise upper bound $U$ and lower bound $L$ of a matrix $Q$, along with the matrix $W$ as inputs.
    \item[$\bullet$] \textbf{Func2} This function, provided in Appendix, calculates the element-wise lower $\hat{L}$ and upper bound $\hat{U}$ of the matrix product $J_i P$, where $J_i$ are the Jacobian matrices as in \eqref{eq:Jacobian}. It takes the element-wise upper $\tilde{U}$ and lower bound $\tilde{L}$ matrices of an arbitrary matrix $P$, and the scalars $a$ and $b$ (bounds on the slope of activation functions) as inputs.
\end{itemize}

Note that once the element-wise bounds on the Jacobian of the NN controller are determined, it is straightforward to calculate the bounds $L_\theta$ and $U_\theta$ in Theorem \ref{thm2} using the \textbf{Func3}, provided in Appendix \ref{app:aux}. Moreover, one can employ bounded activation functions, e.g. $\tanh(\cdot)$ in the output layer of NNs $K_{ij}$ in \eqref{eq:param} to easily calculate the element-wise lower and upper bounds of the NCM $M_\phi(x)$. Likewise, the element-wise lower $L_\phi$ and upper $U_\phi$ bounds of $\dot{M}_\phi(x)$ can be computed via \textbf{Func1}. 

\begin{remark}[Computational complexity of Algorithm \ref{alg}] \label{remark1}
    The  complexity of $\textbf{Func1}(W,L,U)$, where $W \in \mathbb{R}^{m \times n}$, and $L,U \in \mathbb{R}^{n \times p}$ is $\mathcal{O}(mnp)$ and for $\textbf{Func2}(L,U,a,b)$ where $L,U \in \mathbb{R}^{n \times p}$ is $\mathcal{O}(np)$. Finally, for Algorithm \ref{alg}, we have $ \mathcal{O}(Nmnp)$, where $N$ is the total number of layers in the NN controller \eqref{eq:controller}. In  the worst case scenario, i.e., $m=n=p$, our verification method has the computational complexity of $\mathcal{O}(Nn^3)$. On the other hand, both MILP and SMT-based verification methods have significant computational overhead and their complexity grows exponentially with the number of neurons \cite{sun2021learning}.
\end{remark}

{\bf Non-uniform bounded Jacobian of $f(x)$:} After determining the element-wise lower $L$ and upper $U$ bounds inequality \eqref{eq:thm2_main}, our goal becomes computing the positive scalar $\eta = -(c_1 + c_2)$ to leverage Theorem~\ref{thm2} for designing contractivity-enforcing regularizers.
Finding $c_2$ is crucial for this step, and it necessitates a uniformly bounded Jacobian $\partial_x f$ for all $x \in \mathcal{X}$. 
In many real-world applications (e.g., globally Lipschitz-bounded nonlinear systems), the Jacobian of the underlying dynamics exhibits uniform global boundedness (see Section~\ref{sec:experiments}). However, for systems that lack this property, careful gridding in the neighborhood of the desired equilibrium point $x^*$  can be employed. Nevertheless, to provide deterministic guarantees that this uniform bound holds across all $x \in \mathcal{X}$, it is necessary to define an upper bound on the distance between two adjacent gridding samples \cite{sun2021learning}.

Given a Lipschitz continuous function $h: \mathcal{X} \rightarrow \mathbb{R}$ endowed with a Lipschitz constant $L_h$, assume one discretizes  the domain $\mathcal{X}$ such that the distance between any sampled point $x_i$ and its nearest neighbor $x_j$ is less than $\Vert x_j - x_i\Vert < \tau$. If $h (x_i ) < - L_h\tau$ holds for all sampled points $x_i \in \mathcal{X}$, then $h(x) < 0$ holds for all $x \in \mathcal{X}$.
The following proposition provides deterministic guarantees of the existence of $c_2$ and provides an upper bound on the resolution of the gridding samples.

\begin{proposition} \label{prop1}
  Let $L_{\delta_x}$ and $L_M$ be the Lipschitz constants of $\partial_x f$, and $M_\phi(x)$, respectively. Then, $\bar{\lambda}(\text{Sym}[M_\phi(x) \partial_x f(x)])$ has the  Lipschitz constant of $2(S_M L_{\delta_f} + S_{\delta_f} L_M)$, where $S_M$ and $S_{\delta_f}$ satisfy $\Vert M_\phi \Vert_2 \leq S_M$, and $ \Vert \partial_x f(x) \Vert_2 \leq S_{\delta_f}$, respectively.  \hfill $\blacksquare$
\end{proposition}

The detailed proof is given in the Appendix. Likewise, Proposition \ref{prop1} can be utilized to compute the constant $c_1$ in Theorem \ref{thm2}. Moreover, the readers are deferred to \cite[Appendix B.2]{sun2021learning} for more information on the calculations of these Lipschitz constants.  

{\bf Control and NCM design:} Based on Theorem \ref{thm2}, and Algorithm \ref{alg}, we define the following optimization problem to simultaneously train the NN controller \eqref{eq:controller} and NCM $M_\phi(x)$:
    \begin{align} \label{eq:training_problem}
        \min_{\theta, \phi}  \ell_{1}(x^*) + \nu \ell_{2}(\eta, L, U) \;, 
    \end{align}
    with 
    \begin{align*}
        &\ell_1(x^*) = ||f(x^*) + g u_\theta (x^*)|| \\
        &\ell_2(\eta, L, U) = [ 2U_{ii}+ \eta \\ 
        & \qquad \qquad \qquad + \sum_{j\neq i} \max(|L_{ij}+L_{ji}|, |U_{ij}+U_{ji}|)]_{+}\;,
    \end{align*}
    where $[x]_{+} = \max\{0, x\}$ operates element-wise. The regularizer  $\ell_1$ assigns the equilibrium condition, the regularizer $\ell_2$ enforces the contractivity, rendering the equilibrium point exponentially stable, and $\nu$ trades off both regularizers. If both losses converge to zero, we have a valid certificate NCM for the corresponding state-feedback controller policy $u_\theta(x)$. 

\section{Experiments} 
\label{sec:experiments}
In this section, we discuss two numerical examples to validate the efficacy of our method. 
In Section \ref{sec:simple_pendulum}, the objective is to stabilize a standard pendulum and an inverted pendulum.
In Section \ref{sec:example2}, we consider a pre-stabilized non-linear system with  an LQR controller and then, we leverage our method to design an NN controller to expand the region of attraction. 
\subsection{Exponential set-point stabilization}
\begin{figure}
    \centering
    \includegraphics[width = 0.8\linewidth]{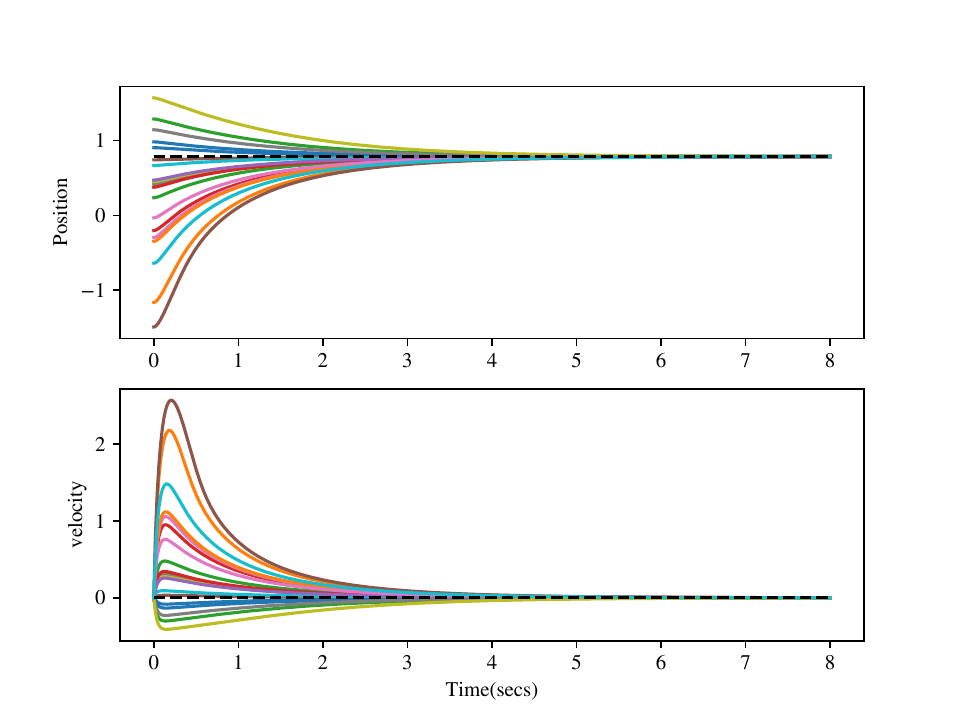}
    \caption{Closed-loop response for 20 different initial conditions demonstrating stabilization at $(\pi/4,0)$.}
    \label{fig:simple_pendulum}
  \end{figure}
\textbf{Standard pendulum:} \label{sec:simple_pendulum}
Consider the following dynamics 
\begin{equation} \label{eq:pendulum}
  \dot{x}=\left[\begin{array}{c}
x_2 \\
\frac{-m g l \sin (x_1)-0.1 x_2+u}{m l^2}
\end{array}\right]  \;,
\end{equation}
where $x_1$ and $x_2$ are the positions and the angular velocity, respectively. Moreover, $g = 9.81$, $m = 0.15$, and $l = 0.5$. 
In this experiment, our goal is to design a NN controller \eqref{eq:controller} such that the closed-loop dynamics are contractive and have the equilibrium point $(x_1^\star,x_2^\star) = (\pi/4,0)$. We choose CCM $M = I$, and for the NN controller, we choose a three-layered network with 32 neurons in the hidden layer, and the activation function \eqref{eq:activation} with $a = 0.3$, and $b = 1.0$. In this case, the Jacobian of the dynamics is
  \begin{align*}
      \partial_x f (x) = \begin{bmatrix}
          0 & 1 \\ -\frac{g}{l} \cos(x_1) & -\frac{0.1}{ml^2}
      \end{bmatrix},
  \end{align*}
  and we can uniformly bound its eigenvalues as  $c_2 = \sup_{x \in \mathbb{R}^n} \bar{\lambda}(\text{Sym}[\partial_x f]) = 20.45$. We trained the NN controller using the loss $\ell = \ell_1 +\ell_2 $ and in this case, the optimal value of $\ell$ was zero.
  This is also evident from the evolution of the closed-loop trajectories starting from 20 arbitrary initial conditions depicted in Fig. \ref{fig:simple_pendulum}, where all the trajectories converge to the desired equilibrium point.\footnote{Our code is available at \href{https://github.com/DecodEPFL/Neural-Exponential-Stabilization}{https://github.com/DecodEPFL/Neural-Exponential-Stabilization}}  
  
\begin{figure}
    \centering
\includegraphics[width = 0.8\linewidth]{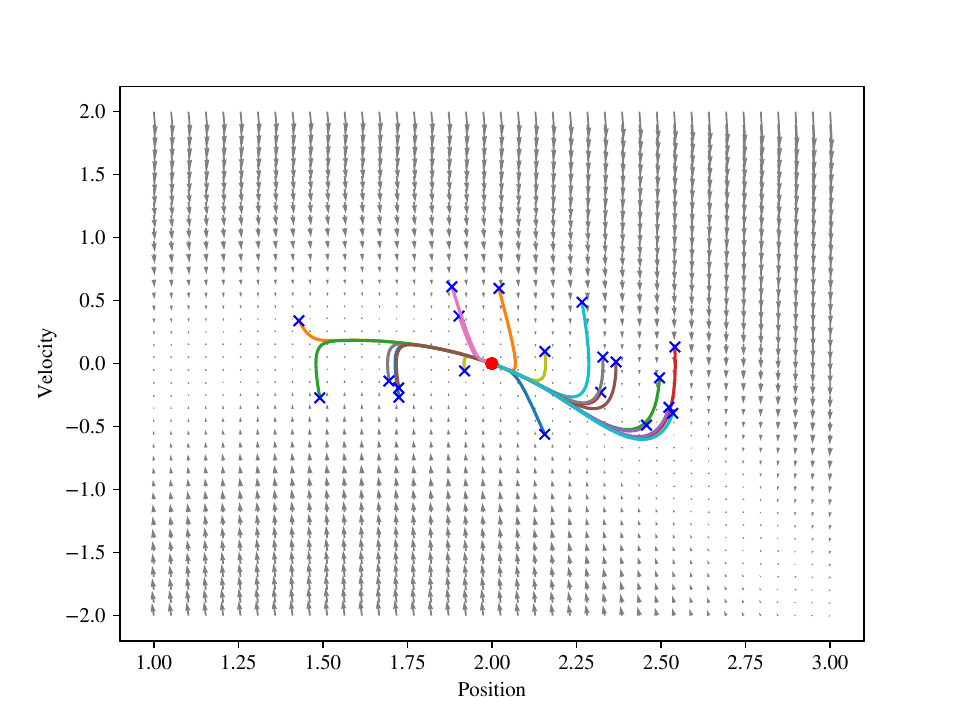}
    \caption{The phase-portrait of the closed-loop system with the trained NN controller. All the trajectories initialized in the neighborhood of the desired equilibrium point are converging exponentially fast.}
    \label{fig:phase_portray_inverted_pendulum}
\end{figure}

\textbf{Inverted pendulum:}
One can easily obtain the dynamics of an inverted pendulum from \eqref{eq:pendulum} by changing the sign in front of $\frac{g}{l}$.
We repeated the same procedure to design the controller to render the closed-loop contractive with the desired equilibrium  point $(x_1^\star,x_2^\star) = (2.0,0.0)$. The evolution of the closed-loop trajectories provided in Fig. \ref{fig:inverted_
pendulum_traj} demonstrate set-point stabilization to the desired equilibrium point, and Fig. \ref{fig:phase_portray_inverted_pendulum} illustrates the phase-portrait. Also, in this case, we obtained a zero training loss.  We selected a contraction rate of $\rho =0.5$ and $a = 0.2, b = 1.0$ in \eqref{eq:activation}. We tuned the learning rate to \texttt{1.0e-3} in both experiments.

\begin{figure}
    \centering
    \includegraphics[width = 0.8\linewidth]{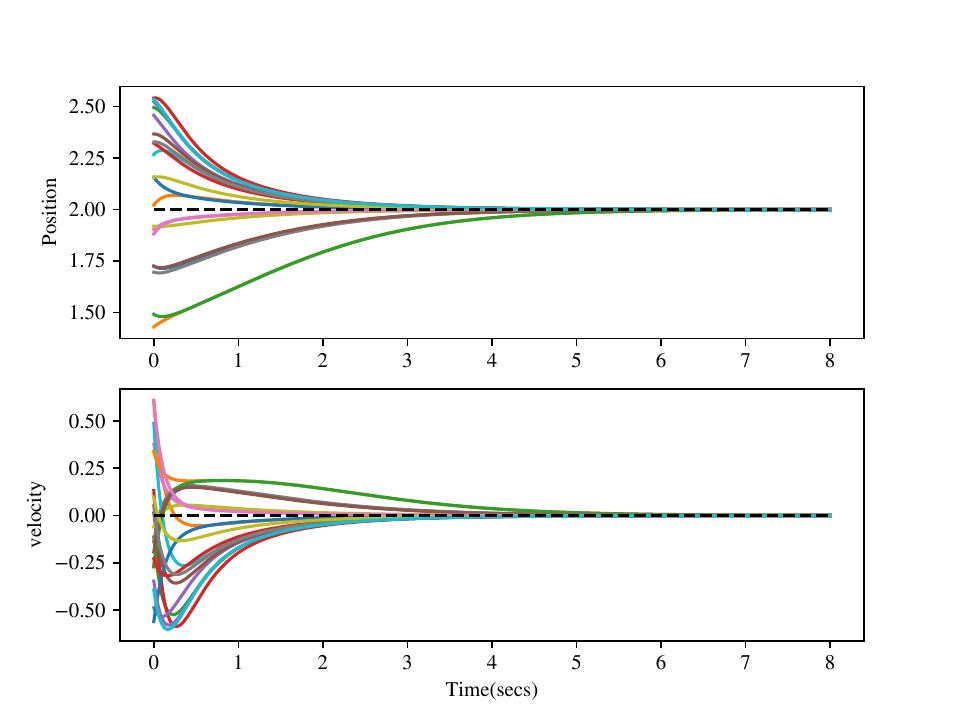}
    \caption{Evolution of closed-loop trajectories of inverted pendulum stabilized at $(2.0,0)$ for 20 arbitrary initial conditions.}
    \label{fig:inverted_
pendulum_traj}
\end{figure}

  \subsection{Enhancing the region of attraction} \label{sec:example2}

As discussed in prior works \cite{manchester2017control, andrieu2008uniting} and references therein, researchers have investigated the challenge of unifying locally optimal and globally stabilizing controllers. This task is particularly difficult within the Lyapunov framework because the set of control Lyapunov functions for a system is non-convex. However, the CCM framework offers a straightforward approach to achieving this goal.
Let us illustrate  with an example taken from \cite{andrieu2008uniting,manchester2017control}, where state $x = [x_1, x_2, x_3]^\top$  and the system follows the dynamics \eqref{eq:plant} with
\begin{equation} \label{eq:sys_example2}
f(x)=\left[\begin{array}{c}
-x_1+x_3 \\
x_1^2-x_2-2 x_1 x_3+x_3 \\
-x_2
\end{array}\right],  g=\left[\begin{array}{l}
0 \\
0 \\
1
\end{array}\right].
\end{equation}
We first solve the linear quadratic regulator (LQR) problem
for the system linearized at the origin with cost function
$\int_0^\infty (x^\top x+r u^2) dt$, where $r = 1$, obtaining a solution $P = P^\top > 0$ 
of the algebraic Riccati equation and the locally optimal
controller $u = -r^{-1}B^\top P x$. 

Next, we trained an NN controller for 2000 epochs to achieve a desired equilibrium point of $(x_1^\star,x_2^\star,x_3^\star) = (0,0,0)$ for the pre-stabilized closed-loop system. The NN controller architecture consisted of a single hidden layer with 64 neurons and utilized the activation function \eqref{eq:activation} with  $a=0.3$ and $b=1.0$. 
For the sake of simplicity, we choose NCM $M_\phi(x)=I$. 

\begin{figure}
    \centering
    \includegraphics[width=\linewidth]{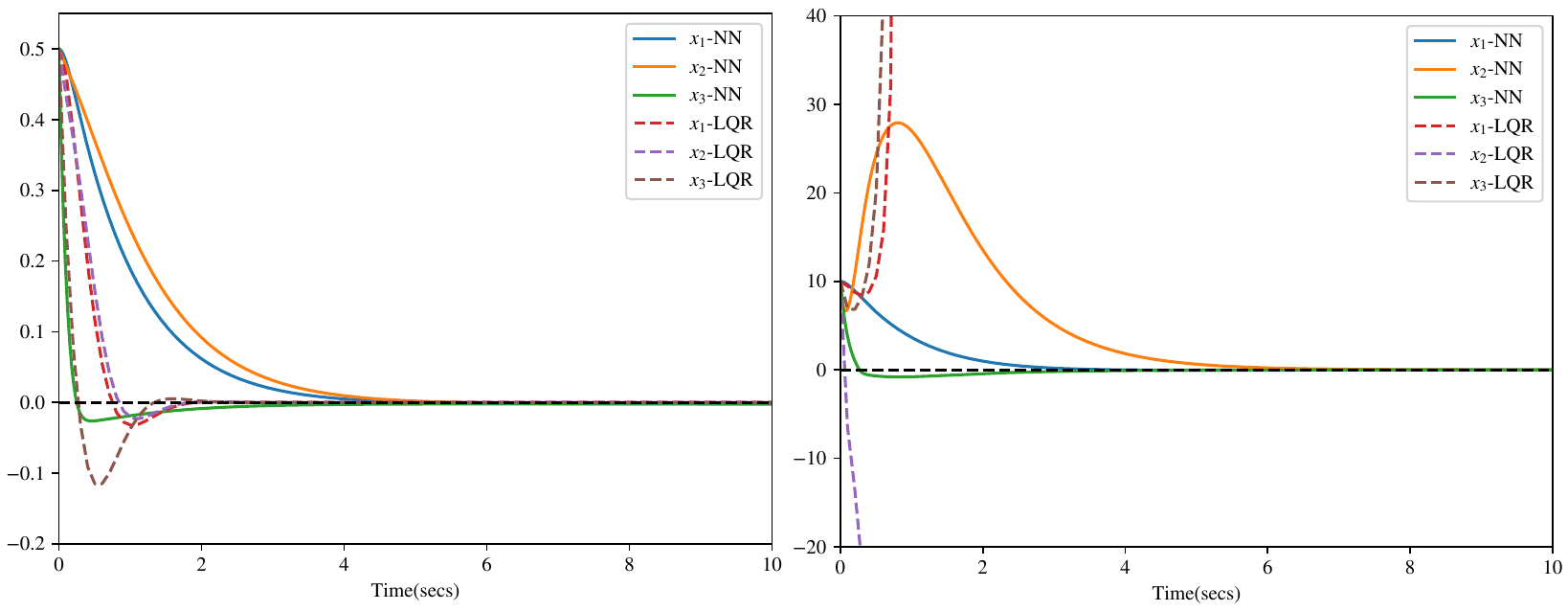}
    \caption{Response of nonlinear system \eqref{eq:sys_example2} with NCM and LQR control to initial state $x(0) = [0.5, 0.5, 0.5]^\top$ (left) and $x(0) = [10, 10, 10]^\top$ (right). This exhibits the locally optimal and increased region of attraction of the NCM controller.}
    \label{fig:ian_manchester_stable}
\end{figure}

Figure \ref{fig:ian_manchester_stable} demonstrates that for small initial conditions, the performance of the NN controller is nearly identical to that of the LQR controller.  
Additionally, with a contraction rate of $\rho$ and $M_\phi(x)$, the NN control law (refer to \eqref{eq:controller}) approximates a basic linear feedback on the error $x - x^\star$.
In contrast, the LQR controller fails to achieve stability for larger initial conditions, while the NN controller successfully stabilizes the system. The simulations under LQR control shown in the right panel of Figure \ref{fig:ian_manchester_stable} diverge rapidly after approximately $1$ second. Conversely, the NN control law demonstrates its efficacy by achieving stability even for a large initial state of $x_0 = [10.0, 10.0, 10.0]^\top$.

\section{Concluding Remarks} \label{sec:conclusions}

For decades, automatic certificate synthesis in control theory has remained elusive. Traditional methods like LP-based synthesis and sums-of-squares programming addressed limitations to some extent, but are not scalable.
In contrast, recently proposed neural certificates offer a promising solution for nonlinear controller synthesis. 

This paper details a novel co-synthesis approach for an NN controller and its NCM certificate for exponential stabilization of a class of nonlinear systems.
Specifically, we transformed an intractable NCM computation (an infinite-dimensional SDP) into a tractable condition based on some element-wise upper and lower bounds of some matrix-valued functions. Our proposed algorithm efficiently computes these bounds. We leveraged this to design contraction-enforcing regularizers for equilibrium-assignment problems and enhancing region of attraction.
Furthermore, under zero training loss, this training scheme provides us with a formal verification of the NN controller. Finally, we introduce a novel NCM parametrization ensuring positive definiteness and a PDE constraint satisfaction regardless of trainable parameters. The future efforts will be devoted to generalizing our approach to other classes of nonlinear systems.

\section*{APPENDIX}
\subsection{An auxiliary Lemma}
The following lemma provides a sufficient condition for the negative definiteness of an arbitrary matrix-valued function. The proof is based on the Gershgorin disk theorem and is omitted here.  
\begin{lemma} \label{lemma1}
    Suppose $L \in \mathbb{R}^{n \times n}$ and $U \in \mathbb{R}^{n \times n}$ are element-wise lower and upper bounds of a matrix-valued function $Y:\mathcal{X} \mapsto \mathbb{R}^{n \times n}$, respectively. Then $Y(x)$ is uniformly negative definite, i.e.  $Y(x) + Y(x)^\top \prec -\rho$, where $\rho > 0$ and $x \in \mathcal{X}$, if the following condition is satisfied
    \begin{align*}
       2U_{ii}+\rho \le - \sum_{j\neq i} \max(|L_{ij}+L_{ji}|, |U_{ij}+U_{ji}|) \;.
    \end{align*}
\end{lemma}



\subsection{Proof of Theorem 2}
Our goal is to satisfy \eqref{eq:cont_ineq} for all $x \in \mathcal{X}$, i.e.,
     \begin{align*}
 \dot{M}_\phi(x) + \text{Sym}[M_\phi(x)  (F(x,u) + g u_\theta(x)) ] \preceq -2\rho M_\phi(x)\;.
  \end{align*}
One can write the above inequality as\footnote{For simplicity, we will often omit the dependence on $x$ when it's clear from the context.}
  \begin{gather}
    \dot{M}_\phi +  \text{Sym}\left[ M_\phi \partial_x f + M_\phi g \partial_x u_\theta  \right] \preceq -2 \rho M_\phi \nonumber \\
  \dot{M}_\phi +  \text{Sym}\left[ M_\phi g \partial_x u_\theta \right]  \preceq -2\rho M_\phi - \bar{\lambda}( \text{Sym}\left[M_\phi \partial_x f \right]) \nonumber  \\
  \dot{M}_\phi +   \text{Sym}\left[ M_\phi g \partial_x u_\theta \right]  \preceq \eta \;,  \label{eq:lMi_proof}        
  \end{gather}
  where  $$\eta :=  - 2\rho \sup_{s \in \mathcal{X}} \bar{\lambda}(M_\phi(s))  
    - \sup_{s \in \mathcal{X}}\bar{\lambda}\left(\text{Sym}[M_\phi(s) \partial_s f(s)]  \right)\;.$$ 
Since  $x\in \mathcal{X}$, where $\mathcal{X}$ is compact, the uniform bound on the dynamic eigenvalues $\eta$ always exists.  
Next, we compute the upper and lower bounds for  
$$ \text{Sym}[M_\phi g \partial_x u_\theta ] =  \text{Sym}[ M_\phi g J_NW_N\ldots J_1W_1] \,,$$ where we substituted the Jacobian of the controller \eqref{eq:controller} from \eqref{eq:Jacobian}. First, using Algorithm \ref{alg}, we compute element-wise bounds $L_{\partial_x u}, U_{\partial_x u}$ for the Jacobian $J_NW_N\ldots J_1W_1 $. Second, we employ \textbf{Func1} in Appendix \ref{app:aux} to compute intermediate bounds $[L_\text{int}, U_\text{int}] = \textbf{Func1}(L,U,g)$. Third, leveraging \textbf{Func3} in Appendix \ref{app:aux}, we calculate $[L_\theta, U_\theta]= \textbf{Func3}(L_M,U_M, L_{\text{int}},U_{\text{int}})$, where 
   $L_\theta \leq  M_\phi g \partial_x u_\theta \leq U_\theta \;.$
Likewise, one can determine the element-wise bounds on $\dot{M}_\phi$, that is $L_\phi \leq  \dot{M}_\phi \leq U_\phi$ using the same procedure. 
Finally, defining  $L := L_\theta + L_\phi$, and $U := U_\theta + U_\phi$ the element-wise bounds on the LHS of LMI \eqref{eq:lMi_proof}
and in light of Lemma  \ref{lemma1}, we obtain the result in Theorem \ref{thm2}.
    \hfill $\blacksquare$

\subsection{Proof of Proposition \ref{prop:ccm}}
We start the proof by considering a single entry of $M_{\phi,ij}(x)$ as 
\begin{align} \label{eq:jacobian_proof}
    M_{\phi,ij}(x) = \sum_{k} \Gamma_{\phi, ik}^\top(x) \Gamma_{\phi, kj}(x) \;,
\end{align}
where the parametrization of $\Gamma_{\phi,(\cdot,\cdot)}$ is given in \eqref{eq:param}. Then the Jacobian of \eqref{eq:jacobian_proof} can be computed as  
\begin{equation}\label{eq:jacobian_gamma}
    \sum_{k} \partial_x (\Gamma_{\phi,ik}^\top \Gamma_{\phi,k\ell})  
= \sum_k \Gamma_{\phi, ik}^\top \partial_x \Gamma_{\phi, kl} + \partial_x \Gamma_{\phi, ik}^\top \Gamma_{\phi, kl} \;, 
 \end{equation}
 where 
 \begin{align*} 
  \partial_x \Gamma_{\phi, \star} =\frac{\partial K_{\star}}{\partial \mathrm{input}} \sum_{i=1}^r \frac{\partial \beta_{\ell,\star}}{\partial \mathrm{input}} {v}_\ell=\sum_{\ell=1}^r \alpha_{\ell, \star}(x) {v}_\ell \, ,
\end{align*}
for some scalar functions $\alpha_{\ell,\star}: \mathcal{X} \rightarrow \mathbb{R}, \, \ell=1, \ldots, r$.  Moreover, $\frac{\partial K_\star}{\partial \text{input}}, \frac{\partial \beta_{\ell,\star}}{\partial \text{input}}$ represent the gradient of $K(\cdot)$ and $\beta_{\ell,\star}(\cdot)$ with respect to their input, respectively. Here $\star$ denotes arbitrary indices of a matrix. Substituting $\partial_x \Gamma_{\phi, \star}$ in \eqref{eq:jacobian_gamma}, we obtain 
\begin{equation}
    \label{eq:full_Exp}
    \sum_k \left[ \Gamma_{\phi, ik}^\top \sum_{\ell=1}^r \alpha_{\ell,kl}(x) {v}_\ell  + \sum_{\ell=1}^r \alpha_{\ell,ik}(x) {v}_\ell \Gamma_{\phi, kl} \right] \;.
\end{equation} 
Note that all the state-dependent functions, i.e., $\Gamma_{\phi, \star}$, and $\alpha_{\ell,\star}$ are scalars. Thus, we can write \eqref{eq:full_Exp} as 
\begin{align*} \label{eq:full_Exp2}
    \sum_k \left[  \sum_{\ell=1}^r \Gamma_{\phi, ik}^\top \alpha_{\ell,kl}(x) {v}_\ell  + \sum_{\ell=1}^r \Gamma_{\phi, kl} \alpha_{\ell,ik}(x)   {v}_\ell\right]
\end{align*}
which is equivalent to 
\begin{align*} 
    \sum_k \left[   \sum_{\ell=1}^r \left( \Gamma_{\phi, ik}^\top \alpha_{\ell,kl}(x)  +  \Gamma_{\phi, kl} \alpha_{\ell,ik}(x) \right)   {v}_\ell\right] \;.
\end{align*}
Finally, we obtain $\sum_k \sum_{\ell=1}^r \zeta_{\phi,k}(x)   {v}_\ell \;,$
where $\zeta_\phi(x) = \left( \Gamma_{\phi, ik}^\top \alpha_{\ell,kl}(x)  +  \Gamma_{\phi, kl} \alpha_{\ell,ik}(x) \right)$. Since $v_\ell$ are the basis of  $\textbf{Ker}(g)$, the product $\partial_x \left( M_{\phi,ij}^\top(x) \right) g$ is zero. Thus, satisfying the PDE \eqref{eq:pde} by design. \hfill $\blacksquare$ 
\subsection{Proof of Proposition \ref{prop1}}

We start the proof by recalling that any two symmetric matrices $A,B \in \mathbb{R}^n$ satisfy the following inequality~\cite{sun2021learning}   
    $$|\bar{\lambda}(A)-\bar{\lambda}(B)| \leq \Vert  A-B \Vert _2 \;.$$
    Moreover, consider two arbitrary matrix-valued functions $A:\mathbb{R}^n \mapsto \mathbb{R}^{k \times l}$ and $B:\mathbb{R}^n \mapsto \mathbb{R}^{l \times s}$ endowed with Lipschitz constants $L_A$ and $L_B$ with respect to the 2-norm. Then, for any $x,y \in \mathbb{R}^n$, one has 
    $$\Vert A(x)B(x)-A(y)B(y) \Vert_2 \leq (S_A L_B + S_B L_A)\Vert  x - y \Vert_2,$$ 
    where $\Vert A(x)\Vert_2 \leq S_A$ and $\Vert B(x) \Vert_2 \leq S_B$ \cite[Lemma 6]{sun2021learning}. To conclude the proof, we employ these inequalities with the matrix-valued functions $M_\phi(x)$ and $ \partial_x f(x)$. Thus, the following hold for all $x,\tilde{x} \in \mathcal{X} \times \mathcal{X}$ 
    \begin{align*}
        |\bar{\lambda}(&\text{Sym}[M_\phi \partial_x f(x)]) - \bar{\lambda}(\text{Sym}[M_\phi \partial_{\tilde{x}} f(\tilde{x})])| \\
        &\leq 2 \Vert M_\phi \partial_x f(x) - M_\phi \partial_{\tilde{x}} f(\tilde{x}) \Vert_2 \leq c_2 \Vert x - \tilde{x} \Vert_2 \;,
    \end{align*}
 where $c_2 : =2 (S_M L_{\delta_f} + S_{\delta_f} L_M)$.  The same arguments can be followed to compute $c_1$ in Theorem \ref{thm2}.     \hfill $\blacksquare$
\subsection{Auxiliary Functions} \label{app:aux}
\begin{itemize}
  \item \textbf{Func1:}
  \begin{itemize}
  \item input: $L, U, W$, output: $\tilde{L}, \tilde{U}$
  \item the procedure is as follows:
    Since $[WQ]_{ij}=\sum_k [W]_{ik}[Q]_{kj}=\sum_{k^+} [W]_{ik^+}[Q]_{k^+j}+\sum_{k^-} [W]_{i{k^-}}[Q]_{{k^-}j}$, where $k^+$ denotes the index of $k$ in $[W]_{ik}$ such that $[W]_{ik}\ge 0$, and  $k^-$ denotes the index of $k$ in $[W]_{ik}$, such that $[W]_{ik}<0$. Then, we have
\begin{align*}
  [  \tilde{L}]_{ij}&= \sum_{k^+} [W]_{ik^+}[L]_{k^+j}+\sum_{k^-} [W]_{i{k^-}}[U]_{{k^-}j}\\
  [  \tilde{U}]_{ij}&= \sum_{k^+} [W]_{ik^+}[U]_{k^+j}+\sum_{k^-} [W]_{i{k^-}}[L]_{{k^-}j} \;.
\end{align*}
\end{itemize}

\item \textbf{Func2:}
  \begin{itemize}
  \item input: $\tilde{L}, \tilde{U}, a, b$, output: $\hat{L}, \hat{U}$
  \item the calculation is given as below:
Since $[JP]_{ij}= [J]_{i}[P]_{ij}$, and $[\tilde{L}]_{ij}\le [P]_{ij}\le [\tilde{U}]_{ij}$, and $[J]_{i}>0$, we obtain
$
[\tilde{L}]_{ij} [J]_{i}\le   [JP]_{ij} \le [J]_{i}[\tilde{U}]_{ij} \;.
$
Then, we have
\begin{itemize}
\item if $ [\tilde{U}]_{ij} \ge [\tilde{L}]_{ij}\ge 0$, then $
    [\hat{L}]_{ij}= a [\tilde{L}]_{ij}, 
    [\hat{U}]_{ij}= b [\tilde{U}]_{ij}$
\item if $ [\tilde{U}]_{ij} \ge 0 \ge [\tilde{L}]_{ij}$, then
    $
    [\hat{L}]_{ij}= b [\tilde{L}]_{ij}, 
    [\hat{U}]_{ij}= b [\tilde{U}]_{ij} $
\item if $0\ge [\tilde{U}]_{ij} \ge [\tilde{L}]_{ij}$, then
$
    [\hat{L}]_{ij}= b [\tilde{L}]_{ij}, \quad
    [\hat{U}]_{ij}= a [\tilde{U}]_{ij} \;.
$
\end{itemize}
\end{itemize}

\item {\bf Func3:} This function calculates the element-wise lower $\bar{L}$ and upper $\bar{U}$ bounds of the product  $W P$.
\begin{itemize}
    \item input: $L_W, U_W, L_P$ and, $U_P$, 
    \item output: $\bar{L}$, and $\bar{U}$
    \item Since $[WP]_{ij} = \sum_{k = 1}^n [W]_{ik}[P]_{kj}$, to find the minimal and maximal of element-wise entries of $W  P$, each term $[W]_{ik}[P]_{kj}$ in the sum should be smallest possible or maximal possible, respectively. Hence, we focus only on a product $[W]_{ik}[P]_{kj}$, where $[L_W]_{ik} \leq [W]_{ik} \leq [U_W]_{ik}$, and  $[L_P]_{kj} \leq [P]_{kj} \leq [U_P]_{kj}$. Then, we have 
    \begin{itemize}
        \item if $[U_W]_{ik} \geq [L_W]_{ik} \geq 0$, and $ [U_P]_{kj}\geq [L_P]_{kj} \geq 0$, then 
          \begin{align*}
    [\bar{L}]_{ij}^k= [L_W]_{ik} [L_P]_{kj}, 
    [\bar{U}]_{ij}^k= [U_W]_{ik} [U_P]_{kj} 
  \end{align*}  
  \item if $0\ge [U_W]_{ik} \ge [L_W]_{ik}$, and $0 \geq [U_P]_{kj}\geq [L_P]_{kj}$, then
        \begin{align*}
    [\bar{L}]_{ij}^k = [U_W]_{ik} [U_P]_{kj},
    [\bar{U}]_{ij}^k = [L_W]_{ik} [L_P]_{kj} 
  \end{align*}
  \item else
          \begin{align*}
    [\bar{L}]_{ij}^k &= \min \{ [L_W]_{ik} [U_P]_{kj}, [U_W]_{ik} [L_P]_{kj}\} \\
    [\bar{U}]_{ij}^k &= \max \{ [U_W]_{ik} [U_P]_{kj}, [L_W]_{ik}[L_P]_{kj}  \} 
  \end{align*}
Finally, the matrices $\bar{L}$, and $\bar{U}$ can be computed as 
       \begin{align*}
    [\bar{L}]_{ij} =  \sum_{k} [\bar{L}]_{ij}^k, \quad 
    [\bar{U}]_{ij}^k =  \sum_{k} [\bar{U}]_{ij}^k 
  \end{align*}
    \end{itemize}
\end{itemize}
\end{itemize}
\bibliography{bibliography}

\begin{thebibliography}{10}
\providecommand{\url}[1]{#1}
\csname url@samestyle\endcsname
\providecommand{\newblock}{\relax}
\providecommand{\bibinfo}[2]{#2}
\providecommand{\BIBentrySTDinterwordspacing}{\spaceskip=0pt\relax}
\providecommand{\BIBentryALTinterwordstretchfactor}{4}
\providecommand{\BIBentryALTinterwordspacing}{\spaceskip=\fontdimen2\font plus
\BIBentryALTinterwordstretchfactor\fontdimen3\font minus
  \fontdimen4\font\relax}
\providecommand{\BIBforeignlanguage}[2]{{%
\expandafter\ifx\csname l@#1\endcsname\relax
\typeout{** WARNING: IEEEtran.bst: No hyphenation pattern has been}%
\typeout{** loaded for the language `#1'. Using the pattern for}%
\typeout{** the default language instead.}%
\else
\language=\csname l@#1\endcsname
\fi
#2}}
\providecommand{\BIBdecl}{\relax}
\BIBdecl

\bibitem{xiao2023barriernet}
W.~Xiao, T.-H. Wang, R.~Hasani, M.~Chahine, A.~Amini, X.~Li, and D.~Rus,
  ``Barriernet: Differentiable control barrier functions for learning of safe
  robot control,'' \emph{IEEE Transactions on Robotics}, 2023.

\bibitem{sun2021learning}
D.~Sun, S.~Jha, and C.~Fan, ``Learning certified control using contraction
  metric,'' in \emph{Conference on Robot Learning}.\hskip 1em plus 0.5em minus
  0.4em\relax PMLR, 2021, pp. 1519--1539.

\bibitem{taylor2019episodic}
A.~J. Taylor, V.~D. Dorobantu, H.~M. Le, Y.~Yue, and A.~D. Ames, ``Episodic
  learning with control {Lyapunov} functions for uncertain robotic systems,''
  in \emph{2019 IEEE/RSJ International Conference on Intelligent Robots and
  Systems (IROS)}.\hskip 1em plus 0.5em minus 0.4em\relax IEEE, 2019, pp.
  6878--6884.

\bibitem{peruffo2021automated}
A.~Peruffo, D.~Ahmed, and A.~Abate, ``Automated and formal synthesis of neural
  barrier certificates for dynamical models,'' in \emph{International
  conference on tools and algorithms for the construction and analysis of
  systems}.\hskip 1em plus 0.5em minus 0.4em\relax Springer, 2021, pp.
  370--388.

\bibitem{tsukamoto2021contraction}
H.~Tsukamoto, S.-J. Chung, and J.-J.~E. Slotine, ``Contraction theory for
  nonlinear stability analysis and learning-based control: A tutorial
  overview,'' \emph{Annual Reviews in Control}, vol.~52, pp. 135--169, 2021.

\bibitem{chang2019neural}
Y.-C. Chang, N.~Roohi, and S.~Gao, ``Neural {Lyapunov} control,''
  \emph{Advances in neural information processing systems}, vol.~32, 2019.

\bibitem{richards2018lyapunov}
S.~M. Richards, F.~Berkenkamp, and A.~Krause, ``The {Lyapunov} neural network:
  {A}daptive stability certification for safe learning of dynamical systems,''
  in \emph{Conference on Robot Learning}.\hskip 1em plus 0.5em minus
  0.4em\relax PMLR, 2018, pp. 466--476.

\bibitem{choi2020reinforcement}
J.~Choi, F.~Castaneda, C.~J. Tomlin, and K.~Sreenath, ``Reinforcement learning
  for safety-critical control under model uncertainty, using control {Lyapunov}
  functions and control barrier functions,'' \emph{arXiv preprint
  arXiv:2004.07584}, 2020.

\bibitem{qin2021learning}
Z.~Qin, K.~Zhang, Y.~Chen, J.~Chen, and C.~Fan, ``Learning safe multi-agent
  control with decentralized neural barrier certificates,'' \emph{arXiv
  preprint arXiv:2101.05436}, 2021.

\bibitem{armenio2019model}
L.~B. Armenio, E.~Terzi, M.~Farina, and R.~Scattolini, ``Model predictive
  control design for dynamical systems learned by echo state networks,''
  \emph{IEEE Control Systems Letters}, vol.~3, no.~4, pp. 1044--1049, 2019.

\bibitem{bonassi2022recurrent}
F.~Bonassi, M.~Farina, J.~Xie, and R.~Scattolini, ``On recurrent neural
  networks for learning-based control: recent results and ideas for future
  developments,'' \emph{Journal of Process Control}, vol. 114, pp. 92--104,
  2022.

\bibitem{terzi2021learning}
E.~Terzi, F.~Bonassi, M.~Farina, and R.~Scattolini, ``Learning model predictive
  control with long short-term memory networks,'' \emph{International Journal
  of Robust and Nonlinear Control}, vol.~31, no.~18, pp. 8877--8896, 2021.

\bibitem{manchester2017control}
I.~R. Manchester and J.-J.~E. Slotine, ``Control contraction metrics: Convex
  and intrinsic criteria for nonlinear feedback design,'' \emph{IEEE
  Transactions on Automatic Control}, vol.~62, no.~6, pp. 3046--3053, 2017.

\bibitem{tan2008stability}
W.~Tan and A.~Packard, ``Stability region analysis using polynomial and
  composite polynomial {Lyapunov} functions and sum-of-squares programming,''
  \emph{IEEE Transactions on Automatic Control}, vol.~53, no.~2, pp. 565--571,
  2008.

\bibitem{dawson2023safe}
C.~Dawson, S.~Gao, and C.~Fan, ``Safe control with learned certificates: A
  survey of neural {Lyapunov}, barrier, and contraction methods for robotics
  and control,'' \emph{IEEE Transactions on Robotics}, 2023.

\bibitem{dai2020counter}
H.~Dai, B.~Landry, M.~Pavone, and R.~Tedrake, ``Counter-example guided
  synthesis of neural network {Lyapunov} functions for piecewise linear
  systems,'' in \emph{2020 59th IEEE Conference on Decision and Control
  (CDC)}.\hskip 1em plus 0.5em minus 0.4em\relax IEEE, 2020, pp. 1274--1281.

\bibitem{schwan2023stability}
R.~Schwan, C.~N. Jones, and D.~Kuhn, ``Stability verification of neural network
  controllers using mixed-integer programming,'' \emph{IEEE Transactions on
  Automatic Control}, 2023.

\bibitem{abate2020formal}
A.~Abate, D.~Ahmed, M.~Giacobbe, and A.~Peruffo, ``Formal synthesis of
  {Lyapunov} neural networks,'' \emph{IEEE Control Systems Letters}, vol.~5,
  no.~3, pp. 773--778, 2020.

\bibitem{chen2021learning}
S.~Chen, M.~Fazlyab, M.~Morari, G.~J. Pappas, and V.~M. Preciado, ``Learning
  {Lyapunov} functions for hybrid systems,'' in \emph{Proceedings of the 24th
  International Conference on Hybrid Systems: Computation and Control}, 2021,
  pp. 1--11.

\bibitem{tsukamoto2020neural}
H.~Tsukamoto and S.-J. Chung, ``Neural contraction metrics for robust
  estimation and control: A convex optimization approach,'' \emph{IEEE Control
  Systems Letters}, vol.~5, no.~1, pp. 211--216, 2020.

\bibitem{lohmiller1998contraction}
W.~Lohmiller and J.-J.~E. Slotine, ``On contraction analysis for non-linear
  systems,'' \emph{Automatica}, vol.~34, no.~6, pp. 683--696, 1998.

\bibitem{pmlr-v202-li23g}
L.~Li, Y.~Duan, G.~Ji, and Y.~Cai, ``Minimum width of leaky-{R}e{LU} neural
  networks for uniform universal approximation,'' in \emph{Proceedings of the
  40th International Conference on Machine Learning}, ser. Proceedings of
  Machine Learning Research, A.~Krause, E.~Brunskill, K.~Cho, B.~Engelhardt,
  S.~Sabato, and J.~Scarlett, Eds., vol. 202.\hskip 1em plus 0.5em minus
  0.4em\relax PMLR, 23--29 Jul 2023, pp. 19\,460--19\,470.

\bibitem{xu2022neural}
L.~Xu, M.~Zakwan, and G.~Ferrari-Trecate, ``Neural energy casimir control for
  {Port-Hamiltonian} systems,'' in \emph{2022 IEEE 61st Conference on Decision
  and Control (CDC)}.\hskip 1em plus 0.5em minus 0.4em\relax IEEE, 2022, pp.
  4053--4058.

\bibitem{andrieu2008uniting}
V.~Andrieu and C.~Prieur, ``Uniting two control {Lyapunov} functions for affine
  systems,'' in \emph{2008 47th IEEE Conference on Decision and Control}.\hskip
  1em plus 0.5em minus 0.4em\relax IEEE, 2008, pp. 622--627.

\end{thebibliography}
\bibliographystyle{IEEEtran}

\end{document}